\documentclass[conference,10pt]{IEEEtran}
\usepackage{cite}
\ifCLASSINFOpdf
\else
   \usepackage[dvipdfmx]{graphicx}
\fi
\usepackage[cmex10]{amsmath}
\usepackage{amssymb}
\usepackage{enumitem}
\usepackage{balance}

\newtheorem{assumption}{Assumption}
\newtheorem{theorem}{Theorem}
\newtheorem{lemma}{Lemma}
\newcommand{\aeq}{\overset{\mathrm{a.s.}}{=}}
\newcommand{\ato}{\overset{\mathrm{a.s.}}{\to}}
\newcommand{\dto}{\overset{\mathrm{d}}{\to}}

\begin{document}
\title{Rigorous Dynamics of Expectation-Propagation-Based Signal Recovery from Unitarily Invariant Measurements}

%\author{Keigo~Takeuchi,~\IEEEmembership{Member,~IEEE}% <-this % stops a space
%\thanks{
%Manuscript received April *, 2015. 
%The author was in part supported by the Grant-in-Aid for Exploratory Research (JSPS KAKENHI Grant Number 15K13987), Japan.
%The material in this paper will be submitted in part to 2016 IEEE 
%International Symposium on Information Theory, Barcelona, Spain, Jul.\ 2016. 
%}
%\thanks{K.~Takeuchi is with the Department of Communication 
%Engineering and Informatics, the University of Electro-Communications, 
%Tokyo 182-8585, Japan (e-mail: ktakeuchi@uec.ac.jp).}% <-this % stops a space
%}
\author{\IEEEauthorblockN{Keigo Takeuchi}
\IEEEauthorblockA{Dept.\ Electrical and Electronic Information Eng., 
Toyohashi University of Technology, 
Aichi 441-8580, Japan}
\IEEEauthorblockA{Email: takeuchi@ee.tut.ac.jp} 
}

\maketitle

\begin{abstract}
This paper investigates sparse signal recovery based on expectation 
propagation (EP) from unitarily invariant measurements. A rigorous analysis 
is presented for the state evolution (SE) of an EP-based message-passing 
algorithm in the large system limit, where both input and output dimensions 
tend to infinity at an identical speed. The main result is the justification 
of an SE formula conjectured by Ma and Ping. 
\end{abstract}

%\begin{IEEEkeywords}
%Massive multiple-input multiple-output (MIMO) systems, uniform linear 
%antenna arrays, antenna spacing, faster-than-Nyquist signaling, 
%large-system analysis. 
%\end{IEEEkeywords}

% \fi
%\IEEEpeerreviewmaketitle

\section{Introduction} \label{sec1}
Consider the $N$-dimensional signal recovery from compressed, linear, 
and noisy measurements $\boldsymbol{y}\in\mathbb{C}^{M}$ ($N\geq M$),  
\begin{equation} \label{system}
\boldsymbol{y} 
= \boldsymbol{A}\boldsymbol{x} + \boldsymbol{w}, 
\quad \boldsymbol{w}\sim\mathcal{CN}(\boldsymbol{0},
\sigma^{2}\boldsymbol{I}_{M}), 
\end{equation}
where $\boldsymbol{x}\in\mathbb{C}^{N}$ and 
$\boldsymbol{A}\in\mathbb{C}^{M\times N}$ denote a sparse signal 
vector and a measurement matrix, respectively. 
The goal is to estimate the unknown signals $\boldsymbol{x}$ from the 
knowledge about the measurement vector $\boldsymbol{y}$ and matrix 
$\boldsymbol{A}$, as well as about the statistics of all random variables. 
Throughout this paper, we postulate the following mild assumptions:  

\begin{assumption} \label{assumption_x}
The signal vector $\boldsymbol{x}$ has independent and identically 
distributed (i.i.d.) zero-mean {\em non}-Gaussian elements\footnote{
We require no additional assumptions for the prior distribution of 
each signal to prove the main theorem, 
whereas it is practically important to postulate some prior distribution 
indicating the sparsity of $\boldsymbol{x}$.  
} with unit variance and finite fourth moments. 
\end{assumption}
\begin{assumption} \label{assumption_A}
The Gram matrix $\boldsymbol{A}^{\mathrm{H}}\boldsymbol{A}$ is unitarily 
invariant~\cite{Tulino04}. Furthermore, the empirical eigenvalue 
distribution of $\boldsymbol{A}\boldsymbol{A}^{\mathrm{H}}$ converges 
almost surely to a deterministic distribution $\rho(\lambda)$ with 
finite fourth moments in the large system limit, where both 
$M$ and $N$ tend to infinity while the compression rate $\delta=M/N\in(0,1]$ 
is kept constant.  
\end{assumption}

For an i.i.d. Gaussian matrix $\boldsymbol{A}$---satisfying 
Assumption~\ref{assumption_A}---the approximate message passing 
(AMP)~\cite{Donoho09} was proved in \cite{Bayati11} to be asymptotically 
Bayes-optimal when the compression rate $\delta$ is larger 
than the so-called belief-propagation (BP) threshold~\cite{Takeuchi15}. 
However, it has been recognized that the original AMP fails to converge for 
non-i.i.d.\ measurement matrices~\cite{Caltagirone14,Rangan14}. To solve 
this limitation, several message-passing algorithms have been proposed 
on the basis of expectation propagation (EP)~\cite{Cespedes14}, 
the expectation-consistent approximation~\cite{Opper05,Kabashima14}, 
or of the turbo principle~\cite{Yuan13,Ma17,Liu16}. These algorithms are 
essentially the same as each other, with the only exception of 
\cite{Kabashima14}. In this paper, they are referred to as EP-based algorithms. 

The main advantage of EP-based algorithms is that they are asymptotically 
Bayes-optimal for unitarily invariant measurement matrices. 
This claim was conjectured in \cite{Ma17}, by proposing state evolution (SE) 
equations of the EP-based algorithms based on two heuristic assumptions, 
and by investigating the properties of the SE equations. 
However, the rigorous justification of the conjecture is 
still open.\footnote{
A similar result~\cite{Rangan16} was posted on the arXiv a few 
months before the first submission of this paper. The main difference between 
the two papers is that we use a probabilistic approach, while the posted paper 
considered a deterministic one based on pseudo-Lipschitz continuity.  
The deterministic approach only provides results averaged over all elements 
of $\boldsymbol{x}$. On the other hand, the probabilistic one allows us to 
obtain results for {\em individual} elements, while only averaged results 
are presented due to space limitation.} 
The purpose of this paper is to prove the conjecture by presenting a 
rigorous derivation of the SE equations. 

\paragraph*{Notation}
The notation $\boldsymbol{o}(1)$ denotes a vector with 
almost surely vanishing Euclidean norm in the large system limit. 
For a matrix $\boldsymbol{M}\in\mathbb{C}^{M\times N}$, 
the singular-value decomposition (SVD) of $\boldsymbol{M}$ is written as 
$\boldsymbol{M}=\boldsymbol{\Phi}_{\boldsymbol{M}}
(\boldsymbol{\Sigma}_{\boldsymbol{M}}, \boldsymbol{O})
\boldsymbol{\Psi}_{\boldsymbol{M}}^{\mathrm{H}}$ for $M\leq N$. The unitary matrix 
$\boldsymbol{\Psi}_{\boldsymbol{M}}=(\boldsymbol{\Psi}_{\boldsymbol{M}}^{\parallel}, 
\boldsymbol{\Psi}_{\boldsymbol{M}}^{\perp})$ is divided into two parts 
that correspond to the non-zero and zero singular values, respectively. 
For $M> N$, we have $\boldsymbol{M}=\boldsymbol{\Phi}_{\boldsymbol{M}}
(\boldsymbol{\Sigma}_{\boldsymbol{M}}^{\mathrm{T}}, \boldsymbol{O})^{\mathrm{T}}
\boldsymbol{\Psi}_{\boldsymbol{M}}^{\mathrm{H}}$ and 
$\boldsymbol{\Phi}_{\boldsymbol{M}}=(\boldsymbol{\Phi}_{\boldsymbol{M}}^{\parallel}, 
\boldsymbol{\Phi}_{\boldsymbol{M}}^{\perp})$. 

When $\boldsymbol{M}$ is full rank, 
the pseudo-inverse of $\boldsymbol{M}$ is denoted by $\boldsymbol{M}^{\dagger}
=\boldsymbol{M}^{\mathrm{H}}(\boldsymbol{M}\boldsymbol{M}^{\mathrm{H}})^{-1}$ 
for $M\leq N$ and by $\boldsymbol{M}^{\dagger}
=(\boldsymbol{M}^{\mathrm{H}}\boldsymbol{M})^{-1}\boldsymbol{M}^{\mathrm{H}}$ 
for $M>N$. Furthermore, 
$\boldsymbol{P}_{\boldsymbol{M}}^{\parallel}=\boldsymbol{M}\boldsymbol{M}^{\dagger}$ 
for $M>N$ represents the orthogonal projection onto the space spanned 
by the columns of $\boldsymbol{M}$, while 
$\boldsymbol{P}_{\boldsymbol{M}}^{\perp}=\boldsymbol{I}_{M}
-\boldsymbol{P}_{\boldsymbol{M}}^{\parallel}$ denotes the projection onto the 
orthogonal complement. 

\section{Expectation Propagation}
We start with an EP-based algorithm~\cite{Cespedes14,Ma17}, which is 
message passing between two modules---called modules A and B. 
Module~A computes the extrinsic mean 
$\boldsymbol{x}_{\mathrm{A}\to \mathrm{B}}^{t}$ and 
covariance $v_{\mathrm{A}\to \mathrm{B}}^{t}\boldsymbol{I}_{N}$ of $\boldsymbol{x}$ 
in iteration~$t$, 
\begin{equation} \label{module_A_mean}
\boldsymbol{x}_{\mathrm{A}\to \mathrm{B}}^{t} 
= \boldsymbol{x}_{\mathrm{B}\to \mathrm{A}}^{t} 
+ \gamma(v_{\mathrm{B}\to\mathrm{A}}^{t})\boldsymbol{W}^{t} 
(\boldsymbol{y} - \boldsymbol{A}\boldsymbol{x}_{\mathrm{B}\to \mathrm{A}}^{t}), 
\end{equation}
\begin{equation} \label{module_A_var}
v_{\mathrm{A}\to \mathrm{B}}^{t} 
=\gamma(v_{\mathrm{B}\to\mathrm{A}}^{t}) - v_{\mathrm{B}\to\mathrm{A}}^{t}
\equiv \varphi_{\mathrm{A}\to \mathrm{B}}(v_{\mathrm{B}\to\mathrm{A}}^{t}), 
\end{equation}
where $\boldsymbol{x}_{\mathrm{B}\to \mathrm{A}}^{t}$ and 
$v_{\mathrm{B}\to \mathrm{A}}^{t}\boldsymbol{I}_{N}$ denote prior   
mean and covariance of $\boldsymbol{x}$ provided from module~B, while 
$\boldsymbol{x}_{\mathrm{B}\to\mathrm{A}}^{0}=\boldsymbol{0}$ and 
$v_{\mathrm{B}\to\mathrm{A}}^{0}=1$ are used in the initial iteration.   
In (\ref{module_A_mean}), the linear minimum mean-square error (LMMSE) 
filter $\boldsymbol{W}^{t}$ is given by  
\begin{equation}
\boldsymbol{W}^{t} 
= \boldsymbol{A}^{\mathrm{H}}
\left(
 \sigma^{2}\boldsymbol{I}_{M} + v_{\mathrm{B}\to \mathrm{A}}^{t}\boldsymbol{A}
 \boldsymbol{A}^{\mathrm{H}}
\right)^{-1}. 
\end{equation}
Furthermore, the function  
$\gamma(v_{\mathrm{B}\to\mathrm{A}}^{t})\equiv\gamma_{t}$ is defined as 
\begin{equation} \label{gamma}  
\frac{1}{\gamma(v_{\mathrm{B}\to\mathrm{A}}^{t})} 
=\lim_{M=\delta N\to\infty}\frac{\mathrm{Tr}
(\boldsymbol{W}^{t}\boldsymbol{A})}{N}
\aeq\int\frac{\delta\lambda d\rho(\lambda)}
{\sigma^{2}+v_{\mathrm{B}\to\mathrm{A}}^{t}\lambda}, 
\end{equation}
due to Assumption~\ref{assumption_A}. 
As proved in Section~\ref{sec4}, $\gamma_{t}$ eliminates 
dependencies between estimation errors in the two modules. 

On the other hand, module~B computes the posterior mean 
$\tilde{\eta}_{t}(\boldsymbol{x}_{\mathrm{A}\to\mathrm{B}}^{t})
=\mathbb{E}[\boldsymbol{x}|\boldsymbol{x}_{\mathrm{A}\to\mathrm{B}}^{t}]$ and 
variance $\mathrm{MMSE}(v_{\mathrm{A}\to\mathrm{B}}^{t})
=N^{-1}\mathbb{E}[\|\boldsymbol{x}
-\tilde{\eta}_{t}(\boldsymbol{x}_{\mathrm{A}\to\mathrm{B}}^{t})\|^{2}
|\boldsymbol{x}_{\mathrm{A}\to\mathrm{B}}^{t}]$ 
of $\boldsymbol{x}$, by regarding the message 
$\boldsymbol{x}_{\mathrm{A}\to\mathrm{B}}^{t}$ as 
the additive white Gaussian noise (AWGN) observation of $\boldsymbol{x}$, 
\begin{equation} \label{AWGN}
\boldsymbol{x}_{\mathrm{A}\to\mathrm{B}}^{t} 
= \boldsymbol{x} + \boldsymbol{\omega}^{t}, 
\quad \boldsymbol{\omega}^{t}\sim\mathcal{CN}(\boldsymbol{0},
v_{\mathrm{A}\to\mathrm{B}}^{t}\boldsymbol{I}_{N}).  
\end{equation}
If a termination condition is satisfied, module~B outputs 
$\tilde{\eta}_{t}(\boldsymbol{x}_{\mathrm{A}\to\mathrm{B}}^{t})$ as an estimate 
of $\boldsymbol{x}$. Otherwise, the extrinsic mean 
$\boldsymbol{x}_{\mathrm{B}\to\mathrm{A}}^{t+1}$ and covariance 
$v_{\mathrm{B}\to\mathrm{A}}^{t+1}\boldsymbol{I}_{N}$ are fed back to module~A. 
\begin{equation} \label{module_B_mean}
\boldsymbol{x}_{\mathrm{B}\to \mathrm{A}}^{t+1} 
= v_{\mathrm{B}\to \mathrm{A}}^{t+1}\left(
 \frac{\tilde{\eta}_{t}(\boldsymbol{x}_{\mathrm{A}\to \mathrm{B}}^{t})}
 {\mathrm{MMSE}(v_{\mathrm{A}\to \mathrm{B}}^{t})}
 - \frac{\boldsymbol{x}_{\mathrm{A}\to \mathrm{B}}^{t}}{v_{\mathrm{A}\to \mathrm{B}}^{t}}
\right)
\equiv \eta_{t}(\boldsymbol{x}_{\mathrm{A}\to \mathrm{B}}^{t}),  
\end{equation}
\begin{equation} \label{module_B_var}
\frac{1}{v_{\mathrm{B}\to \mathrm{A}}^{t+1}} 
= \frac{1}{\mathrm{MMSE}(v_{\mathrm{A}\to \mathrm{B}}^{t})} 
- \frac{1}{v_{\mathrm{A}\to \mathrm{B}}^{t}}
\equiv \frac{1}{\varphi_{\mathrm{B}\to\mathrm{A}}(v_{\mathrm{A}\to \mathrm{B}}^{t})}.  
\end{equation}

The following lemma is used to prove that (\ref{module_B_mean}) eliminates 
dependencies between estimation errors in the two modules.  

\begin{lemma}[Ma and Ping~\cite{Ma17}] \label{lemma_eta}
Let $\boldsymbol{z}^{t}\sim\mathcal{CN}(\boldsymbol{0}, 
v_{\mathrm{A}\to \mathrm{B}}^{t}\boldsymbol{I}_{N})$ denote an 
independent circularly symmetric complex Gaussian random vector with 
covariance $v_{\mathrm{A}\to \mathrm{B}}^{t}\boldsymbol{I}_{N}$. Then, 
\begin{equation}
\lim_{\|\boldsymbol{\epsilon}\|\to0}\mathbb{E}_{\boldsymbol{z}^{t}}\left[
 (\boldsymbol{z}^{t})^{\mathrm{H}}\eta_{t}(\boldsymbol{x}+\boldsymbol{\epsilon}
 +\boldsymbol{z}^{t})
\right] = 0,
\end{equation}
\begin{equation} 
\lim_{\|\boldsymbol{\epsilon}\|\to0}\mathbb{E}_{\boldsymbol{z}^{t}}\left[
 (\boldsymbol{z}^{t})^{*}\tilde{\eta}_{t}(\boldsymbol{x}+\boldsymbol{\epsilon}
 +\boldsymbol{z}^{t})
\right] 
= N\mathrm{MMSE}(v_{\mathrm{A}\to \mathrm{B}}^{t}). 
\end{equation}
\end{lemma}

\section{Main Result}
The following theorem is the main 
result of this paper, which describes the rigorous dynamics of 
the mean-square error (MSE) for the estimate 
$\tilde{\eta}_{t}(\boldsymbol{x}_{\mathrm{A}\to\mathrm{B}}^{t})$ of the 
EP-based algorithm in the large system limit. 

\begin{theorem} \label{theorem1}
Define $\mathrm{mse}_{\mathrm{A}\to\mathrm{B}}^{t} 
= \varphi_{\mathrm{A}\to\mathrm{B}}(\mathrm{mse}_{\mathrm{B}\to\mathrm{A}}^{t})$ and 
$\mathrm{mse}_{\mathrm{B}\to\mathrm{A}}^{t+1} = \varphi_{\mathrm{B}\to\mathrm{A}}
(\mathrm{mse}_{\mathrm{A}\to\mathrm{B}}^{t})$ with 
$\mathrm{mse}_{\mathrm{B}\to\mathrm{A}}^{0}=1$, in which 
$\varphi_{\mathrm{A}\to\mathrm{B}}$ and 
$\varphi_{\mathrm{B}\to\mathrm{A}}$ are defined in (\ref{module_A_var}) and 
(\ref{module_B_var}). Then, the instantaneous MSE 
$\mathrm{mse}_{t} = \lim_{M=\delta N\to\infty}N^{-1}\|\boldsymbol{x}
-\tilde{\eta}_{t}(\boldsymbol{x}_{\mathrm{A}\to\mathrm{B}}^{t})\|^{2}$ for 
the EP-based algorithm converges almost surely to 
$\mathrm{MMSE}(\mathrm{mse}_{\mathrm{A}\to\mathrm{B}}^{t})$ 
in iteration~$t$. 
\end{theorem}

Theorem~\ref{theorem1} was originally conjectured in \cite{Ma17}, and 
implies that the EP-based algorithm predicts the exact dynamics of the 
extrinsic variances in the large system limit. 
The fixed-points (FPs) of the SE equations were proved in \cite{Ma17} to 
correspond to those of an asymptotic energy function that describes the 
Bayes-optimal performance---derived in \cite{Takeda06} via a non-rigorous 
tool in statistical physics. Thus, 
the Bayes-optimal performance derived in \cite{Takeda06} is achievable 
when the SE equations have a unique FP, or equivalently when the compression 
rate $\delta$ is larger than the BP threshold~\cite{Takeuchi15}.    

Let us prove Theorem~\ref{theorem1}. We first formulate the error recursions 
of the EP-based algorithm. 
Let $\boldsymbol{h}_{t}=\boldsymbol{x} 
- \boldsymbol{x}_{\mathrm{A}\to \mathrm{B}}^{t}$ 
and $\boldsymbol{q}_{t}=\boldsymbol{x} 
- \boldsymbol{x}_{\mathrm{B}\to \mathrm{A}}^{t}$ denote the estimation errors 
in modules A and B, respectively. 
From (\ref{system}), (\ref{module_A_mean}), 
(\ref{module_B_mean}), and the SVD 
$\boldsymbol{A}=\boldsymbol{U}(\boldsymbol{\Sigma}, \boldsymbol{O})
\boldsymbol{V}^{\mathrm{H}}$, we obtain the error recursions with 
the initial condition $\boldsymbol{q}_{0}=\boldsymbol{x}$,  
\begin{equation} \label{module_A}
\boldsymbol{m}_{t}
= \boldsymbol{b}_{t} 
- \gamma_{t}\tilde{\boldsymbol{W}}_{t} 
\{(\boldsymbol{\Sigma},\boldsymbol{O})\boldsymbol{b}_{t} 
+ \tilde{\boldsymbol{w}}\},  
\quad \boldsymbol{b}_{t} 
= \boldsymbol{V}^{\mathrm{H}}\boldsymbol{q}_{t},   
\end{equation}
\begin{equation} \label{module_B}
\boldsymbol{q}_{t+1} 
= \boldsymbol{q}_{0} - \eta_{t}(\boldsymbol{q}_{0}-\boldsymbol{h}_{t}), 
\quad \boldsymbol{h}_{t} 
= \boldsymbol{V}\boldsymbol{m}_{t}, 
\end{equation}
with $\tilde{\boldsymbol{w}}=\boldsymbol{U}^{\mathrm{H}}\boldsymbol{w}$. 
In (\ref{module_A}), the linear filter $\tilde{\boldsymbol{W}}_{t}$ is given by 
\begin{equation} \label{Wt}
\tilde{\boldsymbol{W}}_{t}
= (\boldsymbol{\Sigma}, \boldsymbol{O})^{\mathrm{H}}\left(
 \sigma^{2}\boldsymbol{I}_{M} + v_{\mathrm{B}\to \mathrm{A}}^{t}\boldsymbol{\Sigma}^{2}
\right)^{-1}. 
\end{equation}

We next introduce several notations to present a general theorem, of which a 
corollary is Theorem~\ref{theorem1}. 
Let $\boldsymbol{Q}_{t}=(\boldsymbol{q}_{0},\ldots,\boldsymbol{q}_{t-1})
\in\mathbb{C}^{N\times t}$. 
The matrices $\boldsymbol{B}_{t}\in\mathbb{C}^{N\times t}$, 
$\boldsymbol{M}_{t}\in\mathbb{C}^{N\times t}$, 
and $\boldsymbol{H}_{t}\in\mathbb{C}^{N\times t}$ are defined in the same manner.  
The dynamics of the set $\mathcal{X}_{t,t'}
=\{\boldsymbol{Q}_{t+1},\boldsymbol{B}_{t'}, \boldsymbol{M}_{t'}, 
\boldsymbol{H}_{t}| \boldsymbol{B}_{t'}^{\mathrm{H}}\boldsymbol{M}_{t}
=\boldsymbol{Q}_{t'}^{\mathrm{H}}\boldsymbol{H}_{t}, 
\boldsymbol{M}_{t'}=\boldsymbol{G}_{t'}(\boldsymbol{B}_{t'}), 
\boldsymbol{Q}_{t+1}=\boldsymbol{F}_{t}(\boldsymbol{H}_{t},\boldsymbol{q}_{0})
\}$ conditioned on $\Theta=\{\boldsymbol{\Sigma}, \tilde{\boldsymbol{w}}\}$ 
is investigated for $t'=t$ or $t'=t+1$, 
in which the $\tau$th columns of $\boldsymbol{G}_{t'}(\boldsymbol{B}_{t'})$ 
and $\boldsymbol{F}_{t}(\boldsymbol{H}_{t},\boldsymbol{q}_{0})$ are equal to 
the right-hand sides (RHS) on the first equations in (\ref{module_A}) and 
(\ref{module_B}) with $t=\tau$, respectively. The condition 
$\boldsymbol{B}_{t'}^{\mathrm{H}}\boldsymbol{M}_{t}
=\boldsymbol{Q}_{t'}^{\mathrm{H}}\boldsymbol{H}_{t}$ imposes the unitary property 
on $\boldsymbol{V}$, and is obtained from the second 
equations in (\ref{module_A}) and (\ref{module_B}). The set $\mathcal{X}_{t,t}$ 
represents the history of errors in all preceding iterations just before 
updating (\ref{module_A}), while $\mathcal{X}_{t,t+1}$ does 
just before updating (\ref{module_B}). 

We define $\boldsymbol{m}_{t}^{\parallel}=\boldsymbol{P}_{\boldsymbol{M}_{t}}^{\parallel}
\boldsymbol{m}_{t}=\boldsymbol{M}_{t}\boldsymbol{\alpha}_{t}$, 
$\boldsymbol{\alpha}_{t}=\boldsymbol{M}_{t}^{\dagger}\boldsymbol{m}_{t}$, 
and $\boldsymbol{m}_{t}^{\perp}=\boldsymbol{m}_{t}-\boldsymbol{m}_{t}^{\parallel}$. 
See the end of Section~\ref{sec1} for the notations. 
The vectors $\boldsymbol{q}_{t}^{\parallel}$, $\boldsymbol{q}_{t}^{\perp}$, and 
$\boldsymbol{\beta}_{t}=\boldsymbol{Q}_{t}^{\dagger}\boldsymbol{q}_{t}$ are 
defined in the same manner. 
For notational convenience, we define $\boldsymbol{\alpha}_{0}=\boldsymbol{0}$, 
$\boldsymbol{\beta}_{0}=\boldsymbol{0}$, $\boldsymbol{Q}_{0}=\boldsymbol{O}$, 
$\boldsymbol{B}_{0}=\boldsymbol{O}$, $\boldsymbol{M}_{0}=\boldsymbol{O}$, 
$\boldsymbol{H}_{0}=\boldsymbol{O}$, 
$\boldsymbol{M}_{0}^{\dagger}=\boldsymbol{O}$, and 
$\boldsymbol{Q}_{0}^{\dagger}=\boldsymbol{O}$, implying  
$\boldsymbol{P}_{\boldsymbol{M}_{0}}^{\perp}=\boldsymbol{I}_{N}$ and 
$\boldsymbol{P}_{\boldsymbol{Q}_{0}}^{\perp}=\boldsymbol{I}_{N}$. 
\begin{theorem} \label{main_theorem}
For any iteration $\tau=0,1,\ldots$, 
\begin{enumerate}[label=(\alph*)]
\item \label{property1} 
Each element in $\boldsymbol{q}_{\tau+1}$ has finite fourth moments. 
Furthermore, the following limit exists for all $\tau'\leq\tau+1$: 
\begin{equation} \label{qq}
\zeta_{\tau+1,\tau'}
\aeq \lim_{M=\delta N\to\infty}\frac{1}{N}  
\boldsymbol{q}_{\tau'}^{\mathrm{H}}\boldsymbol{q}_{\tau+1}. 
\end{equation}
In particular, the properties 
$\zeta_{\tau+1,\tau+1}=v_{\mathrm{B}\to\mathrm{A}}^{\tau+1}$ and 
$\mathrm{mse}_{\tau}\aeq\mathrm{MMSE}(v_{\mathrm{A}\to\mathrm{B}}^{\tau})$ hold. 
The minimum eigenvalues of $N^{-1}\boldsymbol{M}_{\tau+1}^{\mathrm{H}}
\boldsymbol{M}_{\tau+1}$ and $N^{-1}\boldsymbol{Q}_{\tau+2}^{\mathrm{H}}
\boldsymbol{Q}_{\tau+2}$ are strictly positive in the large system limit. 

\item \label{property2}
Let $\{\boldsymbol{z}_{\tau}\sim\mathcal{CN}(\boldsymbol{0},
\boldsymbol{I}_{N})\}$ denote a sequence of independent standard complex 
Gaussian vectors that are independent of $\boldsymbol{V}$.  
Let $\mu_{\tau} \aeq \lim_{M=\delta N\to\infty}
N^{-1}\|\boldsymbol{q}_{\tau}^{\perp}\|^{2}$ and  
$\nu_{\tau} \aeq \lim_{M=\delta N\to\infty}N^{-1}\|\boldsymbol{m}_{\tau}^{\perp}\|^{2}$, 
and define
\begin{equation} \label{b_tilde}
\tilde{\boldsymbol{b}}_{\tau} 
= \boldsymbol{B}_{\tau}\boldsymbol{\beta}_{\tau} 
+ \boldsymbol{M}_{\tau}\boldsymbol{o}(1) 
+ \boldsymbol{B}_{\tau}\boldsymbol{o}(1) 
+ \mu_{\tau}^{1/2}\boldsymbol{z}_{\tau},  
\end{equation}
\begin{equation} \label{h_tilde}
\tilde{\boldsymbol{h}}_{\tau} 
= \boldsymbol{H}_{\tau}\boldsymbol{\alpha}_{\tau} 
+ \boldsymbol{Q}_{\tau+1}\boldsymbol{o}(1) 
+ \boldsymbol{H}_{\tau}\boldsymbol{o}(1)  
+ \nu_{\tau}^{1/2}\boldsymbol{z}_{\tau}. 
\end{equation}
Then, for any $k\in\mathbb{N}$, all $k$-tuples of the elements in 
$\boldsymbol{b}_{\tau}$ conditioned on $\Theta$ and $\mathcal{X}_{\tau,\tau}$ 
and in $\boldsymbol{h}_{\tau}$ conditioned on $\Theta$ and 
$\mathcal{X}_{\tau,\tau+1}$ converge in distribution to the 
corresponding $k$-tuples for $\tilde{\boldsymbol{b}}_{\tau}$ and 
$\tilde{\boldsymbol{h}}_{\tau}$ in the large system limit. 

\item \label{property3}
Let $\boldsymbol{\omega}\in\mathbb{C}^{N}$ denote any vector that 
is independent of $\boldsymbol{V}$, and satisfies $\lim_{N\to\infty}N^{-1}
\|\boldsymbol{\omega}\|^{2}\aeq1$. 
Suppose that $\boldsymbol{D}$ is any $N\times N$ Hermitian 
matrix such that $\boldsymbol{D}$ depends only on $\boldsymbol{\Sigma}$, 
and that $N^{-1}\mathrm{Tr}(\boldsymbol{D}^{2})$ is almost surely convergent 
as $N\to\infty$. Then, for all $\tau'\leq\tau$ and 
$\tau''\leq \tau+1$ 
\begin{equation} \label{bw} 
\lim_{M=\delta N\to\infty}\frac{1}{N}
\boldsymbol{b}_{\tau}^{\mathrm{H}}\boldsymbol{\omega}
\aeq 0,  
\end{equation}
\begin{equation}
\lim_{M=\delta N\to\infty}\frac{1}{N}
\boldsymbol{b}_{\tau'}^{\mathrm{H}}\boldsymbol{D}\boldsymbol{b}_{\tau}
\aeq \zeta_{\tau,\tau'}\lim_{N\to\infty}\frac{1}{N}\mathrm{Tr}(\boldsymbol{D}), 
\label{bbqq}
\end{equation}
\begin{equation} \label{bm}
\lim_{M=\delta N\to\infty}\frac{1}{N}
\boldsymbol{b}_{\tau'}^{\mathrm{H}}\boldsymbol{m}_{\tau}
\aeq 0, 
\end{equation}
\begin{equation} \label{mmqq}
\lim_{M=\delta N\to\infty}\frac{1}{N}\boldsymbol{m}_{\tau'}^{\mathrm{H}}
\boldsymbol{m}_{\tau}
\aeq \gamma_{\tau,\tau'} - \zeta_{\tau,\tau'},  
\end{equation}
\begin{equation} \label{hhmm}
\lim_{M=\delta N\to\infty}\frac{1}{N}
\boldsymbol{h}_{\tau'}^{\mathrm{H}}\boldsymbol{h}_{\tau} 
\aeq \lim_{M=\delta N\to\infty}\frac{1}{N}\boldsymbol{m}_{\tau'}^{\mathrm{H}}
\boldsymbol{m}_{\tau},
\end{equation} 
\begin{equation} \label{hq} 
\lim_{M=\delta N\to\infty}\frac{1}{N}
\boldsymbol{h}_{\tau}^{\mathrm{H}}\boldsymbol{q}_{\tau''} 
\aeq 0, 
\end{equation}
with 
\begin{equation} \label{gamma_tt}
\gamma_{t,t'}
= \gamma_{t}\gamma_{t'}\int 
\frac{\delta\lambda(\sigma^{2}+\zeta_{t,t'}\lambda)}
{(\sigma^{2}+v_{\mathrm{B}\to \mathrm{A}}^{t}\lambda)
(\sigma^{2}+v_{\mathrm{B}\to \mathrm{A}}^{t'}\lambda)}d\rho(\lambda). 
\end{equation}
\end{enumerate}
\end{theorem}

Theorem~\ref{theorem1} follows immediately from Theorem~\ref{main_theorem}. 
A sketch for the proof of Theorem~\ref{main_theorem} is presented in the 
next section. See \cite{Takeuchi16} for the detailed proof. 

\section{Proof of Theorem~\ref{main_theorem}} \label{sec4}
\subsection{Technical Lemmas}
The proof strategy is based on a conditioning technique used in 
\cite{Bayati11}. A challenging part in the proof is to evaluate the 
distributions of the estimation errors in each iteration conditioned on the 
estimation errors in all preceding iterations. Bayati and 
Montanari~\cite{Bayati11} evaluated the conditional distributions via 
the conditional distribution of the measurement matrix $\boldsymbol{A}$. 
Since the LMMSE filter is used in module~A, the conditional distribution of 
$\boldsymbol{A}$ can be regarded as the posterior distribution of 
$\boldsymbol{A}$ given linear, noiseless, and compressed observations 
of $\boldsymbol{A}$, determined by the estimation errors in all preceding 
iterations. For i.i.d.\ Gaussian measurement matrices, it is well known that 
the posterior distribution is also Gaussian. The proof in \cite{Bayati11} 
heavily relies on this well-known fact. 

The main contribution of this paper is to extend the argument 
in \cite{Bayati11} to the case of the unitary matrix $\boldsymbol{V}$. 
Assumption~\ref{assumption_A} implies that $\boldsymbol{V}$ is independent 
of $\boldsymbol{U}$ and $\boldsymbol{\Sigma}$, and a Haar 
matrix~\cite{Tulino04}---uniformly distributed on the space of all possible 
$N\times N$ unitary matrices. Under coordinate rotations in the row and 
column spaces of $\boldsymbol{V}$, it is possible to show that the linear, 
noiseless, and compressed observation of $\boldsymbol{V}$ is equivalent to 
observing {\em part} of the elements in $\boldsymbol{V}$. Since any Haar 
matrix is bi-unitarily invariant~\cite{Tulino04}, the distribution of 
$\boldsymbol{V}$ after the coordinate rotations is the same as the original 
one. Thus, evaluating the conditional distribution of $\boldsymbol{V}$ 
reduces to analyzing the conditional distribution of a Haar matrix given 
part of its elements. This argument was implicitly used in \cite{Bayati11}.  

Evaluation of this conditional distribution is a technically challenging part 
in this paper, while this part is not required for i.i.d.\ Gaussian 
measurements. We use further coordinate rotations to reveal the 
statistical structure of the conditional Haar matrix. We know that 
a Haar matrix has similar properties to an i.i.d.\ Gaussian matrix as 
$N\to\infty$. In particular, a finite number of linear combinations of 
the elements in a Haar matrix were proved to converge in distribution to 
jointly Gaussian-distributed random variables as 
$N\to\infty$~\cite{Chatterjee08}. 
Note that the classical central limit theorem cannot be used, since the 
elements of a Haar matrix are not independent. Using this asymptotic 
similarity between Haar and i.i.d.\ Gaussian matrices, we arrive at 
the following two lemmas: 

\begin{lemma} \label{lemma_conditional_distribution}
For $t\geq0$, $t'>0$, and $N-t-t'>0$, suppose that 
$\tilde{\boldsymbol{V}}$ is an $(N-t-t')\times(N-t-t')$ Haar matrix 
and independent of $\boldsymbol{V}$, and that both 
$\boldsymbol{Q}_{t'}\in\mathbb{C}^{N\times t'}$ and 
$\boldsymbol{M}_{t}\in\mathbb{C}^{N\times t}$ are full rank for $t>0$. Let 
$\boldsymbol{\epsilon}_{2,0}=\|\boldsymbol{q}_{0}\|^{-2}
\boldsymbol{b}_{0}^{\mathrm{H}}\boldsymbol{m}_{0}\boldsymbol{q}_{0}$, and 
\begin{equation} \label{epsilon}
\boldsymbol{\epsilon}_{1,t} 
= \boldsymbol{\Gamma}_{t}^{\mathrm{H}}
\boldsymbol{H}_{t}^{\mathrm{H}}\boldsymbol{q}_{t}^{\perp}, 
\quad \boldsymbol{\epsilon}_{2,t} 
= \boldsymbol{\Delta}_{t}^{\mathrm{H}}
\boldsymbol{B}_{t+1}^{\mathrm{H}}\boldsymbol{m}_{t}^{\perp}  
\end{equation}
for $t>0$, with $\boldsymbol{\Gamma}_{t} 
= \boldsymbol{M}_{t}^{\dagger}
- \boldsymbol{M}_{t}^{\dagger}\boldsymbol{B}_{t}(\boldsymbol{B}_{t}^{\mathrm{H}}
\boldsymbol{P}_{\boldsymbol{M}_{t}}^{\perp}\boldsymbol{B}_{t})^{-1}
\boldsymbol{B}_{t}^{\mathrm{H}}\boldsymbol{P}_{\boldsymbol{M}_{t}}^{\perp}$ and  
$\boldsymbol{\Delta}_{t} 
= \boldsymbol{Q}_{t+1}^{\dagger} 
- \boldsymbol{Q}_{t+1}^{\dagger}\boldsymbol{H}_{t}
(\boldsymbol{H}_{t}^{\mathrm{H}}\boldsymbol{P}_{\boldsymbol{Q}_{t+1}}^{\perp}
\boldsymbol{H}_{t})^{-1}
\boldsymbol{H}_{t}^{\mathrm{H}}\boldsymbol{P}_{\boldsymbol{Q}_{t+1}}^{\perp}$. 
Then, the following properties hold: 
\begin{equation} \label{bt}
\boldsymbol{b}_{t} 
\sim \boldsymbol{B}_{t}\boldsymbol{\beta}_{t} + \boldsymbol{\epsilon}_{1,t}
+ \boldsymbol{\Phi}_{\boldsymbol{M}_{t}}^{\perp}
\boldsymbol{\Psi}_{\boldsymbol{V}_{01}^{t,t}}^{\perp}
\tilde{\boldsymbol{V}}^{\mathrm{H}}
(\boldsymbol{\Phi}_{\boldsymbol{Q}_{t}}^{\perp}
\boldsymbol{\Phi}_{\boldsymbol{V}_{10}^{t,t}}^{\perp})^{\mathrm{H}}
\boldsymbol{q}_{t} 
\end{equation}
conditioned on $\Theta$ and $\mathcal{X}_{t,t}$ for $t>0$, and for 
all $\tau<t+1$  
\begin{equation} \label{Vq}
\boldsymbol{V}^{\mathrm{H}}\boldsymbol{q}_{\tau} 
\sim \boldsymbol{b}_{\tau} 
+ \boldsymbol{\Phi}_{\boldsymbol{M}_{t}}^{\perp}
\boldsymbol{\Psi}_{\boldsymbol{V}_{01}^{t,t+1}}^{\perp}
\tilde{\boldsymbol{V}}^{\mathrm{H}}
(\boldsymbol{\Phi}_{\boldsymbol{Q}_{t+1}}^{\perp}
\boldsymbol{\Phi}_{\boldsymbol{V}_{10}^{t,t+1}}^{\perp})^{\mathrm{H}}
\boldsymbol{q}_{\tau}, 
\end{equation}
\begin{equation}
\boldsymbol{h}_{t}  
\sim \boldsymbol{H}_{t}\boldsymbol{\alpha}_{t} + \boldsymbol{\epsilon}_{2,t}
+ \boldsymbol{\Phi}_{\boldsymbol{Q}_{t+1}}^{\perp}
\boldsymbol{\Phi}_{\boldsymbol{V}_{10}^{t,t+1}}^{\perp}\tilde{\boldsymbol{V}}
(\boldsymbol{\Phi}_{\boldsymbol{M}_{t}}^{\perp}
\boldsymbol{\Psi}_{\boldsymbol{V}_{01}^{t,t+1}}^{\perp})^{\mathrm{H}}
\boldsymbol{m}_{t}   
\end{equation}
conditioned on $\Theta$ and $\mathcal{X}_{t,t+1}$. 
In these expressions, 
$\boldsymbol{V}_{10}^{t,t'}\in\mathbb{C}^{(N-t')\times t}$ and 
$\boldsymbol{V}_{01}^{t,t'}\in\mathbb{C}^{t'\times(N-t)}$ are given by 
\begin{equation} \label{V10}
\boldsymbol{V}_{10}^{t,t'} 
= (\boldsymbol{\Phi}_{\boldsymbol{Q}_{t'}}^{\perp})^{\mathrm{H}}
\boldsymbol{H}_{t}\boldsymbol{M}_{t}^{\dagger}
\boldsymbol{\Phi}_{\boldsymbol{M}_{t}}^{\parallel},  
\end{equation}
\begin{equation} \label{V01}
\boldsymbol{V}_{01}^{t,t'}
= (\boldsymbol{Q}_{t'}^{\dagger}
\boldsymbol{\Phi}_{\boldsymbol{Q}_{t'}}^{\parallel})^{\mathrm{H}}
\boldsymbol{B}_{t'}^{\mathrm{H}}
\boldsymbol{\Phi}_{\boldsymbol{M}_{t}}^{\perp},  
\end{equation}
with $\boldsymbol{V}_{01}^{0,1}=\boldsymbol{b}_{0}^{\mathrm{H}}
/\|\boldsymbol{q}_{0}\|$. See the end of Section~\ref{sec1} for the 
other notations, as well as  
$\boldsymbol{\Phi}_{\boldsymbol{M}_{0}}^{\perp}=\boldsymbol{I}_{N}$ and 
$\boldsymbol{\Phi}_{\boldsymbol{V}_{10}^{0,1}}^{\perp}=\boldsymbol{I}_{N-1}$. 
\end{lemma}

\begin{lemma} \label{lemma_Haar_CLT}
For $t'>t\geq0$ and $N-t-t'>0$, suppose that 
$\tilde{\boldsymbol{V}}$ is the Haar matrix defined in 
Lemma~\ref{lemma_conditional_distribution}.  
Let $\boldsymbol{a}\in\mathbb{C}^{N-t-t'}$ denote a vector that are independent 
of $\tilde{\boldsymbol{V}}$ and satisfies 
$\lim_{N\to\infty}N^{-1}\|\boldsymbol{a}\|^{2}\aeq1$. 
Suppose that $\boldsymbol{z}\in\mathbb{C}^{N}$ is a vector such that, 
for all $k\in\mathbb{N}$, any $k$-tuple of the elements in 
$\boldsymbol{z}$ follows $\mathcal{CN}(\boldsymbol{0},\boldsymbol{I}_{k})$ 
as $N\to\infty$. 
\begin{itemize}
\item If the minimum eigenvalues of  
$N^{-1}\boldsymbol{M}_{t}^{\mathrm{H}}\boldsymbol{M}_{t}$ and 
$N^{-1}\boldsymbol{B}_{t}^{\mathrm{H}}\boldsymbol{P}_{\boldsymbol{M}_{t}}^{\perp}
\boldsymbol{B}_{t}$ are strictly positive in the large system limit, then 
the convergence in distribution holds   
\begin{equation} \label{CLT_M}
\boldsymbol{\Phi}_{\boldsymbol{M}_{t}}^{\perp}
\boldsymbol{\Psi}_{\boldsymbol{V}_{01}^{t,t}}^{\perp}
\tilde{\boldsymbol{V}}^{\mathrm{H}}\boldsymbol{a} 
\dto \boldsymbol{z} + \boldsymbol{M}_{t}\boldsymbol{o}(1) 
+ \boldsymbol{P}_{\boldsymbol{M}_{t}}^{\perp}
\boldsymbol{B}_{t}\boldsymbol{o}(1)
\end{equation}
conditioned on $\boldsymbol{a}$, $\Theta$, and $\mathcal{X}_{t,t}$ 
in the large system limit. 

\item If the minimum eigenvalues of 
$N^{-1}\boldsymbol{Q}_{t+1}^{\mathrm{H}}\boldsymbol{Q}_{t+1}$ and 
$N^{-1}\boldsymbol{H}_{t}^{\mathrm{H}}\boldsymbol{P}_{\boldsymbol{Q}_{t+1}}^{\perp}
\boldsymbol{H}_{t}$ are strictly positive in the large system limit, then 
the convergence in distribution holds 
\begin{equation} \label{CLT_Q} 
\boldsymbol{\Phi}_{\boldsymbol{Q}_{t+1}}^{\perp}
\boldsymbol{\Phi}_{\boldsymbol{V}_{10}^{t,t+1}}^{\perp}
\tilde{\boldsymbol{V}}\boldsymbol{a} 
\dto \boldsymbol{z} + \boldsymbol{Q}_{t+1}\boldsymbol{o}(1) 
+ \boldsymbol{P}_{\boldsymbol{Q}_{t+1}}^{\perp}
\boldsymbol{H}_{t}\boldsymbol{o}(1) 
\end{equation}
conditioned on $\boldsymbol{a}$, $\Theta$, and $\mathcal{X}_{t,t+1}$ 
in the large system limit. 
\end{itemize}
\end{lemma}

In order to prove Theorem~\ref{main_theorem}, we need the strong law of 
large numbers for the elements of a Haar matrix, which are 
dependent random variables. 
%\begin{theorem}[Etemadi~\cite{Etemadi83}] 
%\label{theorem_SLLN}
%Let $\{X_{i}\}_{i=1}^{\infty}$ denote a sequence of complex random variables 
%with finite second moments, and define $S_{n}=\sum_{i=1}^{n}X_{i}$.
%The strong law of large numbers for 
%$T_{n}=(S_{n}-\mathbb{E}[S_{n}])/n$ holds, i.e.\  
%$\lim_{n\to\infty}T_{n}\aeq 0$, if the following assumption holds:     
%\begin{equation} \label{assumption_var}
%\limsup_{n\to\infty}\frac{1}{n^{\gamma}}\mathbb{V}[S_{n}]<\infty
%\quad \hbox{for some $\gamma<2$.} 
%\end{equation}
%\end{theorem}

\begin{lemma} \label{lemma_Haar_norm} 
Suppose that $\boldsymbol{V}$ is an $N\times N$ Haar matrix. 
Let $\boldsymbol{a}\in\mathbb{C}^{N}$ and $\boldsymbol{b}\in\mathbb{C}^{N}$ 
denote random vectors that are independent of $\boldsymbol{V}$ and satisfy  
$\lim_{N\to\infty}N^{-1}\|\boldsymbol{a}\|^{2}\aeq1$, 
$\lim_{N\to\infty}N^{-1}\|\boldsymbol{b}\|^{2}\aeq1$, and 
$\lim_{N\to\infty}N^{-1}\boldsymbol{b}^{\mathrm{H}}\boldsymbol{a}\aeq C$. 
Furthermore, we define a Hermitian matrix 
$\boldsymbol{D}\in\mathbb{C}^{N\times N}$ such that 
$\boldsymbol{D}$ is independent of $\boldsymbol{V}$, and 
that $N^{-1}\mathrm{Tr}(\boldsymbol{D}^{2})$ is almost surely convergent 
as $N\to\infty$. Then, 
\begin{equation}
\lim_{N\to\infty}\frac{1}{N}\boldsymbol{b}^{\mathrm{H}}\boldsymbol{V}
\boldsymbol{a} 
\aeq 0, 
\end{equation}
\begin{equation}
\lim_{N\to\infty}\frac{1}{N}\boldsymbol{b}^{\mathrm{H}}\boldsymbol{V}^{\mathrm{H}}
\boldsymbol{D}\boldsymbol{V}\boldsymbol{a}
\aeq C\lim_{N\to\infty}\frac{1}{N}\mathrm{Tr}(\boldsymbol{D}). 
\end{equation}
\end{lemma}

\subsection{Sketch of Proof by Induction}
We are ready to prove Theorem~\ref{main_theorem}. The proof is by induction. 
We omit the proof for the case $\tau=0$, and only present a sketch 
of the proof for a general case, because of space limitation. 

We assume that Theorem~\ref{main_theorem} is correct for all 
$\tau< t$, and prove that Theorem~\ref{main_theorem} holds for 
$\tau=t$. Note that we can use Lemma~\ref{lemma_conditional_distribution}, 
since the induction hypothesis~\ref{property1} for $\tau<t$ implies that 
$\boldsymbol{M}_{t}$ and $\boldsymbol{Q}_{t'}$ are full rank for $t'=t$ 
and $t'=t+1$. 

\begin{IEEEproof}[Convergence of $\boldsymbol{b}_{\tau}$ to (\ref{b_tilde}) 
for $\tau=t$]
We first prove $\boldsymbol{\epsilon}_{1,t}\aeq \boldsymbol{o}(1)$ 
in (\ref{bt}), given by (\ref{epsilon}). 
We use the submultiplicative property of the Euclidean norm to obtain  
the upper bound $\|\boldsymbol{\epsilon}_{1,t}\|^{2}
\leq \|N\boldsymbol{\Gamma}_{t}\|^{2}
\|N^{-1}\boldsymbol{H}_{t}^{\mathrm{H}}\boldsymbol{q}_{t}^{\perp}\|^{2}$. 

Let us prove that $N^{-1}\boldsymbol{H}_{t}^{\mathrm{H}}
\boldsymbol{q}_{t}^{\perp}$ converges almost surely to zero in the large 
system limit. By definition, 
\begin{equation}
\frac{1}{N}\boldsymbol{H}_{t}^{\mathrm{H}}
\boldsymbol{q}_{t}^{\perp}
= \frac{\boldsymbol{H}_{t}^{\mathrm{H}}\boldsymbol{q}_{t}}{N} 
- \frac{\boldsymbol{H}_{t}^{\mathrm{H}}\boldsymbol{Q}_{t}}{N}
\left(
 \frac{\boldsymbol{Q}_{t}^{\mathrm{H}}\boldsymbol{Q}_{t}}{N}
\right)^{-1}
\frac{\boldsymbol{Q}_{t}^{\mathrm{H}}\boldsymbol{q}_{t}}{N}.  
\end{equation}
The induction hypothesis~(\ref{hq}) for $\tau<t$ implies that 
$N^{-1}\boldsymbol{H}_{t}^{\mathrm{H}}\boldsymbol{q}_{t}$ and 
$N^{-1}\boldsymbol{H}_{t}^{\mathrm{H}}\boldsymbol{Q}_{t}$ converge almost surely 
to zero. Furthermore, 
the induction hypothesis~\ref{property1} for $\tau<t$ 
implies that $\|(N^{-1}\boldsymbol{Q}_{t}^{\mathrm{H}}\boldsymbol{Q}_{t})^{-1}
N^{-1}\boldsymbol{Q}_{t}^{\mathrm{H}}\boldsymbol{q}_{t}\|$ is bounded. 
Thus, $N^{-1}\boldsymbol{H}_{t}^{\mathrm{H}}\boldsymbol{q}_{t}^{\perp}\ato0$ holds 
in the large system limit. 

In order to complete the proof of 
$\boldsymbol{\epsilon}_{1,t}\aeq \boldsymbol{o}(1)$, 
we need to prove that $\|N\boldsymbol{\Gamma}_{t}\|^{2}$ is bounded. 
The boundedness can be proved in the same manner, although the details are 
omitted. Thus, $\boldsymbol{\epsilon}_{1,t}\aeq \boldsymbol{o}(1)$ holds. 

We next use Lemma~\ref{lemma_Haar_CLT} to evaluate the last term 
on the RHS of (\ref{bt}). It is possible to confirm that the last term 
on the RHS of (\ref{CLT_M}) reduces to 
$\boldsymbol{P}_{\boldsymbol{M}_{t}}^{\perp}\boldsymbol{B}_{t}\boldsymbol{o}(1)
= \boldsymbol{M}_{t}\boldsymbol{o}(1) 
+ \boldsymbol{B}_{t}\boldsymbol{o}(1)$. 
Thus, we use (\ref{bt}) and Lemma~\ref{lemma_Haar_CLT} to find that, for 
all $k\in\mathbb{N}$, any $k$-tuple of the elements in $\boldsymbol{b}_{t}$ 
conditioned on $\Theta$ and $\mathcal{X}_{t, t}$ converges in distribution 
to the corresponding $k$-tuple for (\ref{b_tilde}), 
when $\mu_{t}$ in (\ref{b_tilde}) is defined as 
\begin{equation} \label{mut}
\mu_{t} 
\aeq \lim_{M=\delta N\to\infty}\frac{1}{N}\boldsymbol{q}_{t}^{\mathrm{H}}
\boldsymbol{\Phi}_{\boldsymbol{Q}_{t}}^{\perp}
\boldsymbol{P}_{\boldsymbol{V}_{10}^{t,t}}^{\perp}
(\boldsymbol{\Phi}_{\boldsymbol{Q}_{t}}^{\perp})^{\mathrm{H}}
\boldsymbol{q}_{t}. 
\end{equation}

In order to complete the proof, we shall evaluate (\ref{mut}). 
Since we can show 
$\boldsymbol{\Phi}_{\boldsymbol{Q}_{t}}^{\perp}
\boldsymbol{P}_{\boldsymbol{V}_{10}^{t,t}}^{\perp} 
(\boldsymbol{\Phi}_{\boldsymbol{Q}_{t}}^{\perp})^{\mathrm{H}}
= \boldsymbol{P}_{\boldsymbol{Q}_{t}}^{\perp} 
- \boldsymbol{P}_{\boldsymbol{P}_{\boldsymbol{Q}_{t}}^{\perp}\boldsymbol{H}_{t}}^{\parallel}$, 
we have 
\begin{equation} \label{mut_tmp}
\mu_{t} 
\aeq \lim_{M=\delta N\to\infty}\frac{1}{N}\left(
 \|\boldsymbol{q}_{t}^{\perp}\|^{2} 
 - \boldsymbol{q}_{t}^{\mathrm{H}}
 \boldsymbol{P}_{\boldsymbol{P}_{\boldsymbol{Q}_{t}}^{\perp}\boldsymbol{H}_{t}}^{\parallel}
 \boldsymbol{q}_{t} 
\right). 
\end{equation}
It is possible to prove that the second term converges almost surely to 
zero, by repeating the proof of $\boldsymbol{\epsilon}_{1,t}=\boldsymbol{o}(1)$. 
Thus, we have $\mu_{t} \aeq \lim_{M=\delta N\to\infty}
N^{-1}\|\boldsymbol{q}_{t}^{\perp}\|^{2}$. 
\end{IEEEproof}

\begin{IEEEproof}[Eqs.~(\ref{bw})--(\ref{mmqq}) for $\tau=t$]
We first prove (\ref{bbqq}) for $\tau=t$. 
We use (\ref{bt}), $\boldsymbol{\epsilon}_{1,t}
=\boldsymbol{o}(1)$, and Lemma~\ref{lemma_Haar_norm} to have 
\begin{equation}
\lim_{M=\delta N\to\infty}
\frac{1}{N}\boldsymbol{b}_{\tau'}^{\mathrm{H}}\boldsymbol{D}\boldsymbol{b}_{t}
\aeq \lim_{M=\delta N\to\infty}
\frac{1}{N}\boldsymbol{b}_{\tau'}^{\mathrm{H}}\boldsymbol{D}\boldsymbol{B}_{t}
\boldsymbol{\beta}_{t} 
\end{equation}
conditioned on $\Theta$ and $\mathcal{X}_{t,t}$ for $\tau'<t$. 
Using the induction hypothesis~(\ref{bbqq}) for $\tau<t$,
$\boldsymbol{q}_{t}^{\parallel}=\boldsymbol{Q}_{t}\boldsymbol{\beta}_{t}$, 
and $\boldsymbol{q}_{\tau'}^{\mathrm{H}}\boldsymbol{q}_{t}^{\perp}=0$ yields 
(\ref{bbqq}) for $\tau'<\tau=t$. 

For $\tau'=t$, (\ref{bt}) and Lemma~\ref{lemma_Haar_norm} imply 
\begin{IEEEeqnarray}{rl}
\frac{1}{N}\boldsymbol{b}_{t}^{\mathrm{H}}\boldsymbol{D}\boldsymbol{b}_{t}
\ato& 
\frac{1}{N}\boldsymbol{\beta}_{t}^{\mathrm{H}}\boldsymbol{B}_{t}^{\mathrm{H}}
\boldsymbol{D}\boldsymbol{B}_{t}\boldsymbol{\beta}_{t} 
\nonumber \\  
&+ \frac{\mu_{t}}{N}\mathrm{Tr}\left\{
 \boldsymbol{D}\boldsymbol{\Phi}_{\boldsymbol{M}_{t}}^{\perp}
 \boldsymbol{P}_{\boldsymbol{V}_{01}^{t,t}}^{\perp}
 (\boldsymbol{\Phi}_{\boldsymbol{M}_{t}}^{\perp})^{\mathrm{H}}
\right\} 
\end{IEEEeqnarray}
conditioned on $\Theta$ and $\mathcal{X}_{t,t}$ 
in the large system limit. The induction hypothesis~(\ref{bbqq}) for $\tau<t$ 
implies that the fist term converges almost surely to 
$\lim_{M=\delta N\to\infty}N^{-1}
\|\boldsymbol{q}_{t}^{\parallel}\|^{2}N^{-1}\mathrm{Tr}(\boldsymbol{D})$. 
Furthermore, it is possible to prove that the second term 
converges almost surely to 
$\mu_{t}N^{-1}\mathrm{Tr}(\boldsymbol{D})$  
in the large system limit, since $N^{-1}\|\boldsymbol{D}\|^{2}$ is assumed to 
be bounded as $N\to\infty$. Thus, (\ref{bbqq}) holds. 

To prove (\ref{bw}) for $\tau=t$, we repeat the same proof to obtain  
\begin{equation}
\lim_{M=\delta N\to\infty}\frac{1}{N}
\boldsymbol{b}_{t}^{\mathrm{H}}\boldsymbol{\omega}
\aeq \lim_{M=\delta N\to\infty}\frac{1}{N}
\boldsymbol{\beta}_{t}^{\mathrm{H}}\boldsymbol{B}_{t}^{\mathrm{H}}
\boldsymbol{\omega}
\aeq 0  
\end{equation}
conditioned on $\Theta$ and $\mathcal{X}_{t,t}$, 
where we have used the induction hypothesis~(\ref{bw}) for $\tau<t$. 

Let us prove (\ref{bm}) and (\ref{mmqq}) for $\tau=t$. 
Using (\ref{gamma}), (\ref{Wt}), (\ref{bw}), and (\ref{bbqq}), we obtain 
\begin{equation} \label{bm_tmp}
\frac{\gamma_{t}}{N}
\boldsymbol{b}_{\tau'}^{\mathrm{H}}\tilde{\boldsymbol{W}}_{t}\left\{
 (\boldsymbol{\Sigma}, \boldsymbol{O})\boldsymbol{b}_{t} 
 + \tilde{\boldsymbol{w}}
\right\} 
\ato \frac{1}{N}\boldsymbol{b}_{\tau'}^{\mathrm{H}}\boldsymbol{b}_{t} 
\end{equation}
in the large system limit. From (\ref{module_A}) and (\ref{bm_tmp}), 
(\ref{bm}) holds. 

Similarly, we use (\ref{module_A}), (\ref{bbqq}), and (\ref{bm_tmp}) to obtain 
\begin{IEEEeqnarray}{rl}
\frac{\boldsymbol{m}_{\tau'}^{\mathrm{H}}\boldsymbol{m}_{t}}{N} 
\ato - \zeta_{t,\tau'} 
+& \frac{\gamma_{\tau'}\gamma_{t}}{N}
\left\{
 (\boldsymbol{\Sigma}, \boldsymbol{O})\boldsymbol{b}_{\tau'} 
 + \tilde{\boldsymbol{w}}
\right\}^{\mathrm{H}}
\nonumber \\ 
&\cdot\tilde{\boldsymbol{W}}_{\tau'}^{\mathrm{H}} 
\tilde{\boldsymbol{W}}_{t}\left\{
 (\boldsymbol{\Sigma}, \boldsymbol{O})\boldsymbol{b}_{t} 
 + \tilde{\boldsymbol{w}}
\right\}
\end{IEEEeqnarray}
in the large system limit. Using (\ref{Wt}), (\ref{bw}), (\ref{bbqq}),  
and Assumption~\ref{assumption_A}, we find that the second term reduces 
to (\ref{gamma_tt}) for $t'=\tau'$. Thus, (\ref{mmqq}) holds for $\tau=t$.  
\end{IEEEproof}

\begin{IEEEproof}[Convergence of $\boldsymbol{h}_{\tau}$ to 
(\ref{h_tilde}) for $\tau=t$]
The proof for the convergence of $\boldsymbol{h}_{t}$ is omitted, 
since it is the same as for the 
convergence of $\boldsymbol{b}_{t}$ to (\ref{b_tilde}). 
\end{IEEEproof}

\begin{IEEEproof}[Eq.~(\ref{hhmm}) for $\tau=t$]
The proof of (\ref{hhmm}) for $\tau=t$ is omitted, since 
it is the same as the proof for (\ref{bbqq}) with 
$\boldsymbol{D}=\boldsymbol{I}_{N}$. 
\end{IEEEproof}

\begin{IEEEproof}[Eq.~(\ref{qq}) for $\tau=t$]
We only prove the existence of (\ref{qq}) for $\tau'\leq \tau=t$, since 
the case $\tau'=t+1$ can be proved in the same manner. 
Using (\ref{module_B}) yields 
\begin{equation}
\frac{1}{N}\boldsymbol{q}_{\tau'}^{\mathrm{H}}\boldsymbol{q}_{t+1}
= \frac{\boldsymbol{q}_{\tau'}^{\mathrm{H}}\boldsymbol{q}_{0}}{N} 
- \frac{\boldsymbol{q}_{\tau'}^{\mathrm{H}}\eta_{t}
(\boldsymbol{q}_{0}-\boldsymbol{h}_{t})}{N}. 
\end{equation} 
The induction hypothesis~(\ref{qq}) $\tau<t$ implies that 
the first term is convergent in the large system limit. 

In order to prove the existence of (\ref{qq}) for $\tau'\leq \tau=t$, 
it is sufficient to confirm 
\begin{equation} \label{qq_asym}
\frac{1}{N}\boldsymbol{q}_{\tau'}^{\mathrm{H}}\boldsymbol{q}_{t+1}
\ato \zeta_{0,\tau'} 
- \frac{1}{N}\mathbb{E}_{\mathcal{Z}_{t}}\left[
 \boldsymbol{q}_{\tau'}^{\mathrm{H}}\eta_{t}
 (\boldsymbol{q}_{0}-\boldsymbol{h}_{t}^{\mathrm{G}})
\right]
\end{equation}
in the large system limit. In (\ref{qq_asym}), 
the expectation is over independent standard complex Gaussian vectors 
$\mathcal{Z}_{t}=\{\boldsymbol{z}_{\tau}: \tau=0,\ldots,t\}$. 
Furthermore, $\boldsymbol{h}_{\tau}^{\mathrm{G}}$ is recursively defined as 
\begin{equation} \label{h_G}
\boldsymbol{h}_{\tau}^{\mathrm{G}} 
= \boldsymbol{H}_{\tau}^{\mathrm{G}}\boldsymbol{\alpha}_{\tau} 
+ \boldsymbol{Q}_{\tau+1}\boldsymbol{o}(1) 
+ \boldsymbol{H}_{\tau}\boldsymbol{o}(1)+ \nu_{\tau}^{1/2}\boldsymbol{z}_{\tau},  
\end{equation}
with $\boldsymbol{H}_{\tau}^{\mathrm{G}}=(\boldsymbol{h}_{0}^{\mathrm{G}},\ldots,
\boldsymbol{h}_{\tau-1}^{\mathrm{G}})$. 

We use the strong law of large numbers~\cite[Theorem~6]{Lyons88} 
and the property~\ref{property2} for $\tau=t$ to have 
\begin{equation}
\frac{\boldsymbol{q}_{\tau'}^{\mathrm{H}}\boldsymbol{q}_{t+1} }{N}
\ato \zeta_{0,\tau'} 
- \frac{\mathbb{E}_{\boldsymbol{z}_{t}}[
 \boldsymbol{q}_{\tau'}^{\mathrm{H}}
 \eta_{t}(\boldsymbol{q}_{0}-\tilde{\boldsymbol{h}}_{t})]}{N}
\end{equation}
in the large system limit, 
where $\tilde{\boldsymbol{h}}_{t}$ is given by (\ref{h_tilde}). 
Repeating the same argument in the order $\tau=t-1,\ldots,0$, 
we arrive at (\ref{qq_asym}). Thus, 
(\ref{qq}) exists for $\tau'\leq \tau=t$. 
\end{IEEEproof}

\begin{IEEEproof}[Eq.~(\ref{hq}) for $\tau=t$]
We shall prove (\ref{hq}) for $\tau=t$. From (\ref{module_B}) and 
(\ref{Vq}), we find 
\begin{equation}
\lim_{M=\delta N\to\infty}\frac{1}{N}\boldsymbol{h}_{t}^{\mathrm{H}}
\boldsymbol{q}_{\tau'} 
\aeq \lim_{M=\delta N\to\infty}\frac{1}{N} 
\boldsymbol{m}_{t}^{\mathrm{H}}\boldsymbol{b}_{\tau'}, 
\end{equation}
conditioned on $\Theta$ and $\mathcal{X}_{t,t+1}$ for $\tau'\leq t$, 
which is almost surely equal to zero, 
because of (\ref{bm}) for $\tau=t$. 
Thus, (\ref{hq}) holds for $\tau'\leq\tau=t$. 

We use (\ref{module_B}) and (\ref{hq}) for $\tau'=0$ and $\tau=t$ to have 
\begin{equation} \label{hqt}
\lim_{M=\delta N\to\infty}
\frac{1}{N}\boldsymbol{h}_{t}^{\mathrm{H}}\boldsymbol{q}_{t+1} 
\aeq -\lim_{M=\delta N\to\infty}\frac{1}{N}\boldsymbol{h}_{t}^{\mathrm{H}}
\eta_{t}(\boldsymbol{q}_{0}-\boldsymbol{h}_{t}) 
\end{equation}
for $\tau'=t+1$. It is possible to prove 
\begin{equation}
\frac{1}{N}\boldsymbol{h}_{t}^{\mathrm{H}}\boldsymbol{q}_{t+1} 
\ato -\frac{1}{N}\mathbb{E}_{\mathcal{Z}_{t}}\left[
  (\boldsymbol{h}_{t}^{\mathrm{G}})^{\mathrm{H}}
  \eta_{t}(\boldsymbol{q}_{0}-\boldsymbol{h}_{t}^{\mathrm{G}})
\right], \label{hqt_tmp}
\end{equation}
in the large system limit, by repeating the proof of (\ref{qq_asym}). 

Let us prove that the RHS of (\ref{hqt_tmp}) is equal to zero. 
From (\ref{gamma}), (\ref{mmqq}), (\ref{hhmm}), (\ref{gamma_tt}), 
and the induction hypothesis $\zeta_{t,t}=v_{\mathrm{B}\to\mathrm{A}}^{t}$, 
we find that 
the random vector $\boldsymbol{h}_{t}^{\mathrm{G}}$ induced from the randomness 
of $\mathcal{Z}_{t}$ has i.i.d.\ proper complex Gaussian elements with 
vanishing mean in the large system limit and variance 
$v_{\mathrm{A}\to\mathrm{B}}^{t}$, given by (\ref{module_A_var}). 
We use Lemma~\ref{lemma_eta} to find that the RHS of (\ref{hqt_tmp}) 
is equal to zero. Thus, (\ref{hq}) holds for $\tau=t$. 
\end{IEEEproof}

\balance 

\begin{IEEEproof}[Property~\ref{property1} for $\tau=t$]
The proof for the existence of the fourth moments is omitted. 
See \cite[Lemma~1(g)]{Bayati11} for evaluating the minimum 
eigenvalues of $N^{-1}\boldsymbol{M}_{\tau+1}^{\mathrm{H}}
\boldsymbol{M}_{\tau+1}$ and $N^{-1}\boldsymbol{Q}_{\tau+2}^{\mathrm{H}}
\boldsymbol{Q}_{\tau+2}$ for $\tau=t$. 

We repeat the proof of (\ref{qq_asym}) to obtain  
\begin{equation} \label{mse0}
\mathrm{mse}_{t} 
\aeq \lim_{M=\delta N\to\infty}\frac{1}{N}\mathbb{E}\left[
 \|\boldsymbol{q}_{0} - \tilde{\eta}_{t}
(\boldsymbol{q}_{0}-\boldsymbol{h}_{t}^{\mathrm{G}})\|^{2} 
\right],
\end{equation}
which reduces to $\mathrm{MMSE}(v_{\mathrm{A}\to\mathrm{B}}^{t})$.  

Let us prove $\zeta_{t+1,t+1}=v_{\mathrm{B}\to\mathrm{A}}^{t+1}$. 
Applying (\ref{module_B_mean}) to (\ref{module_B}), and using 
(\ref{module_B_var}), we have   
\begin{equation}
\boldsymbol{q}_{t+1} 
= \frac{v_{\mathrm{B}\to \mathrm{A}}^{t+1}\{\boldsymbol{q}_{0} 
- \tilde{\eta}_{t}(\boldsymbol{q}_{0}-\boldsymbol{h}_{t})\}}
{\mathrm{MMSE}(v_{\mathrm{A}\to \mathrm{B}}^{t})}
- \frac{v_{\mathrm{B}\to \mathrm{A}}^{t+1}}{v_{\mathrm{A}\to \mathrm{B}}^{t}} 
\boldsymbol{h}_{t}. 
\end{equation}
We use Lemma~\ref{lemma_eta}, (\ref{module_B_var}), and (\ref{hq}) 
to evaluate (\ref{qq}) as  
\begin{equation}
\zeta_{t+1,t+1} 
\aeq \frac{(v_{\mathrm{B}\to \mathrm{A}}^{t+1})^{2}}
{\mathrm{MMSE}(v_{\mathrm{A}\to \mathrm{B}}^{t})} 
- \frac{(v_{\mathrm{B}\to \mathrm{A}}^{t+1})^{2}}{v_{\mathrm{A}\to \mathrm{B}}^{t}}
= v_{\mathrm{B}\to \mathrm{A}}^{t+1}.  
\end{equation}
Thus, property~\ref{property1} holds. 
\end{IEEEproof}

\section*{Acknowledgment}
The author was in part supported by the Grant-in-Aid 
for Exploratory Research (JSPS KAKENHI Grant Number 15K13987), Japan.

%\ifCLASSOPTIONcaptionsoff
%  \newpage
%\fi



\bibliographystyle{IEEEtran}
\bibliography{IEEEabrv,kt-isit2017}
\end{document}